\documentclass[a4paper,english]{rnti}
\usepackage[english]{babel}
\usepackage[T1]{fontenc}
\usepackage{url}
\usepackage{graphicx}
\usepackage{fancyhdr}

\titrecourt{Automatic Integration Issues of Tabular Data for On-Line Analysis Processing}
\nomcourt{Y. Yang et al.}
\titre{Automatic Integration Issues of Tabular Data for On-Line Analysis Processing}
\auteur{Yuzhao Yang\affil{1},
       J\'er\^ome Darmont\affil{2},
        Franck Ravat\affil{1}, 
         Olivier Teste\affil{1}}
\affiliation{
    \affil{1} Institut de Recherche en Informatique de Toulouse -IRIT CNRS (UMR 5505)- 
    \\Universit\'e de Toulouse, 118, Route de Narbonne, 31069 Toulouse Cedex 9, France
    \\\{Yuzhao.Yang, Franck.Ravat, Olivier.Teste\}@irit.fr, 
    \\\\
    \affil{2}ERIC UR 3083, Universit\'e de Lyon, Lyon 2 
    \\5 avenue Pierre Mend\`es France, F69676 Bron Cedex, France
    \\ Jerome.Darmont@univ-lyon2.fr
}
    
\resume{%
Companies and individuals produce numerous tabular data. The objective of this position paper is to draw up the challenges posed by the automatic integration of data in the form of tables so that they can be cross-analyzed. We provide a first automatic solution for the integration of such tabular data to allow On-Line Analysis Processing. To fulfill this task, features of tabular data is analyzed and the challenges of automatic multidimensional schema generation should be addressed. Hence, we propose a typology of tabular data and a automatic process based on different steps to integrate one or more data sources.
}

\summary{%
Les entreprises et les particuliers produisent de nombreuses donn\'ees tabulaires. L'objectif de cet article est de mettre en avant les d\'efis pos\'es par l'int\'egration automatique des donn\'ees tabulaires pour effectuer des analyses en ligne (OLAP). Pour remplir cette t\^ache, les caract\'eristiques des donn\'ees tabulaires doivent \^etre analys\'ees et les d\'efis de la g\'en\'eration automatique de sch\'emas multidimensionnels doivent \^etre relev\'es. Par cons\'equent, nous proposons une typologie des donn\'ees tabulaires et une d\'emarche automatique bas\'ee sur diff\'erentes \'etapes pour int\'egrer aussi bien une source de donn\'ees que plusieurs.
}

\begin{document}

\section{Introduction}

Business Intelligence (BI) plays an important role in numerous companies and administrations to efficiently support decision making processes. With the current digitization trend, even small companies and organizations can exploit a large number of data every day and the rise of open data make various data even more accessible. Nevertheless, the implementation of a BI project needs to be realized by people who have the professional knowledge and deep skills in BI technologies such as data warehousing and data visualization. Such projects are also usually expensive and time-consuming. As a result, it is necessary to find a solution to automate the BI process to allow small enterprises, organizations and even individuals without deep technical expertise to easily analyze data. Up to now, there is no platform that achieves this goal. 

In current BI systems, data are extracted and stored in a data warehouse to allow On-Line Analysis Processing (OLAP) and visual data rendering. Thus, automating the data warehousing process is crucial to allow non-specialist to exploit such approaches. There exist various forms of data, but most of the data in small enterprises and organizations, as well as most of the open data are in tabular form from spreadsheet software. Although there are commercial BI tools allowing the exploitation of tabular data such as Excel, Qlikview or Tableau, none of them automates the multidimensional analysis of tabular data.

Unlike relational databases that are well-structured and where we can easily retrieve keys, cardinalities and other table metadata, or XML files which sometimes provide schemas such as Document Type Definitions (DTDs), spreadsheet tabular data do not directly provide such information. Moreover, there are different types and structures of tabular data, which makes it difficult to automatically identify labels, values and aggregates. Without input from the user, it is hard to understand the semantics of some data and to choose the appropriate measures for OLAP. Different tables also need to be integrated by matching and mapping. Facing all these requirements and problems, we propose a possible solution in this paper. Table identification is the first significant step we introduce to reach our goal. To carry out this step, we also define an extended typology of tabular data to classify different table types.

The remainder of this paper is organized as follows. In Section 2, we review the related works about automatic multidimensional schema generation and typologies of tabular data, respectively. In Section 3, we define our typology of tabular data. In Section 4, we present our tabular data integration solution, and discuss related perspectives. Finally, in Section 5, we conclude this paper and hint at future research.

\section{Related works}

\subsection{Automatic multidimensional schema generation}

OLAP (On-Line Analytical Processing) systems allow analysts to improve decision-making process by consulting and analyzing aggregated data. A multidimensional schema organizes data according to analysis subjects (facts) associated to analysis axes (dimensions). Each fact is composed of measures. Each dimension contains one or several analysis viewpoints (hierarchies). Each hierarchy contains various data granularities of analysis data. 

There are different approaches for the generation of multidimensional schemas  \citep{Romero09}. There are top-down (also called demand-driven) approaches that start from user requirements and generate the schema to satisfy these requirements manually or automatically. Conversely there are bottom-up approaches that are mostly automatic or semi-automatic solutions to generate the schema from the data source. Moreover, there are hybrid approaches taking both user requirements and the data source into account. Our work focuses on the bottom-up approaches because the users we target do not necessarily know or anticipate precise requirements. 
\begin{sloppypar}
Most of the bottom-up processes are designed for relational databases. \citep{Phipps02} introduce an automatic data warehouse design approach whose input is an Entity-Relationship (ER) schema. They consider numeric fields as measures, the more numeric fields a table contains, the more likely it is to be a fact, so they create for each table a multidimensional schema by taking the numeric fields in the table as measures. Then, they identify the datetime type data to create temporal dimensions. The fields that are not key, numeric or date of the table become dimensions. Finally, they complete dimensions by verifying if the tables connecting with the fact are in the many side of the relationship and complete hierarchies by verifying the many-to-one relationship of the remaining tables. However, they use queries to evaluate what candidate schemas best meet the user’s needs, which is done manually. Thus, the process is not fully automated.
\end{sloppypar}
\citep{Song07} also propose a semi-automatic method to generate star schemas from ER diagrams. The main difference with the approach by \citep{Phipps02} is that the fact table is chosen by calculating the Connection Topology Value (CTV) of each entity, which is a composite function of the topology value of direct and indirect many-to-one relationships. The selected entities are those that have a CTV higher than a threshold calculated by the mean and standard deviation of the CTVs. In the end, the user must check the redundancy of the time dimension, merge the related dimensions and rename the elements. 

There are also methods for other data types. \citep{Golfarelli01} propose a semi-automatic method to construct a conceptual data warehouse schema from XML sources. In their approach, schema design is mainly based on DTDs, as they define the elements and attributes and describe the relationships between elements. They simplify the DTDs and create a DTD graph which represent the structure of the XML. Then, based on the facts chosen by the designer and the DTD graph, they build an attribute tree representing the conceptual schema. 

Eventually, data mining methods are also used to automate multidimensional schema generation. \citep{Usman11} propose to use hierarchical clustering. Their architecture includes a data mining layer to preprocess the dataset and cluster data, and an automatic schema generation layer to generate the multidimensional schema. The solution is implemented and tested with the dataset $ForestCoverType$, but the authors did not specify the similarity metric they use nor the schema generation algorithm. 

In summary, existing multidimensional schema generation approaches are designed for specific data sources, such as relational databases and XML documents, and they are not fully automated. There is no such approach specially designed for tabular data. Moreover, the automatic generation of multidimensional schema for tabular data is specific and there are still many challenges to face compared to other data types (Section 1).

\subsection{Typologies of tabular data}

There is a lot of research concerning tabular data and especially web tables. There are different table classification methods. \citep{Wang02} and \citep{Crestan11} divide web tables into two types: genuine or relational knowledge tables, and non-genuine or layout tables, based on whether tables contain relational knowledge or used for only grouping contents for easy viewing. 
\begin{sloppypar}
We are only interested in genuine tables/relational knowledge tables because non-genuine/layout tables are only used for the navigational or formatting purpose on the web. \citep{Crestan11} further divide relational knowledge tables into listing tables, attribute/value tables, matrix tables, enumeration tables and form tables. This classification method is refined and completed by \citep{Lautert13}, who add new types of relational knowledge tables: concise tables, nested tables, multivalued tables and split tables.
\end{sloppypar}
\citep{Yoshida01} define nine table types according to whether attributes are arranged vertically or horizontally, whether there is a header and how the header is positioned. These types are presented by a graph. \citep{Milosevic16} give a classification of structural tables through dimensionality: one-dimensional tables, two dimensional tables and multidimensional tables that can also be typed into super-row tables and multi-tables. \citep{Chen13} also introduce the concept of hierarchical spreadsheet which contains the header with multi-levels.

All these classifications describe tabular data by different aspects such as the direction of the arrangement of data, the characteristics of cells, the dimensionality of tables, the functionality of tables, etc. However, when considering all table types, none of these classifications are complete. For example, the case of a table whose header is distributed in two rows due to the limit of the space is considered in the classification of \citep{Yoshida01}, but is not considered by the other people, while the matrix table is not considered in their classification.

\section{Augmented tabular data classification}

Data are presented differently in various types of tables. How to extract them in different types of tables is thus difficult. Therefore, identifying the type of table is important and requires a complete classification of tabular data. Thus, we propose a new typology of tabular data (FIG. 1) that fuse and complements the classifications from Section 2.2 with respect to three aspects: the structure of tables, the content of cells and headers which are respectively in gray, blue and green in the figure.

\begin{figure}[h]
\begin{center}
 \includegraphics[width=12cm]{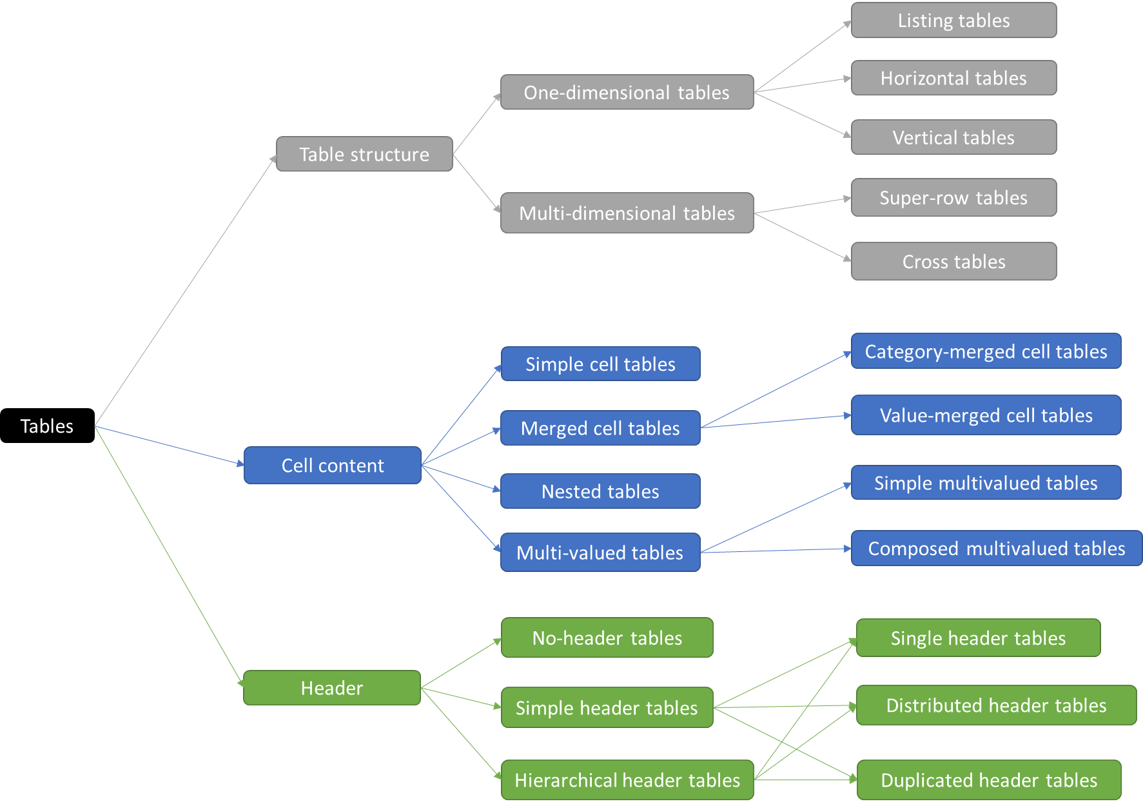}
 \caption{Typology of tabular data} \label{Typology}
\end{center}
\end{figure}

\subsection{Table structure}

The structure of tables aspect concerns the dimensionality of data and the arrangement of rows and columns. With respect to the dimensionality of tables, these latter are classified into one-dimensional tables and multidimensional tables, which also bear sub-types.

\begin{figure}[h]
\begin{center}
 \includegraphics[width=12cm]{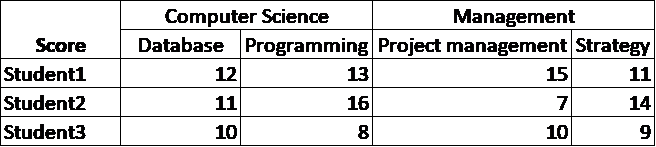}
 \caption{Cross table, Hierarchical header table} \label{CrossHie}
\end{center}
\end{figure}

\begin{enumerate}
\item $One$-$dimensional$ $tables$ have only one dimension.
\begin{itemize}
\item $Listing$ $tables$ list values belonging to a same attribute.
\item In $horizontal$ $tables$, data are arranged horizontally, $i.e$., each column represents one attribute, each row represents a tuple containing values of different attributes.
\item In $vertical$ $tables$, data are arranged vertically, $i.e$.  each row represents one attribute, each column represents a tuple containing values of different attributes.
\end{itemize}

\item $Multidimensional$ $tables$ have multiples dimensions.
\begin{itemize}
\item $Super$-$row$ $tables$ contain multiple dimensions arranged into different level of one row/column, as classified and exampled by Milosevic et al. (2016)
\item $Cross$ $tables$ are usually two-dimensional tables where there is one dimension arranged in column and one other arranged in row (Table in FIG. 2). The value of each cell is determined by the dimensions of the column and the row in which it is located. In the cases where there are more than two dimensions, we can have cross tables for different dimensions.
\end{itemize}

\end{enumerate}

\subsection{Cell content}
This concerns the characteristics of cell content, as classified and exampled by \citep{Lautert13}.
\begin{enumerate}
\item	$Simple$ $cell$ $tables$ contain only “normal” cells without any of the exceptions listed below.
\item	$Merged$ $cell$ $tables$ contain cells that are gathered or merged into one cell.
\begin{itemize}
\item	$Category$-$merged$ $cell$ $tables$ contain cells merged together because they belong to the same category.
\item	$Value$-$merged$ $cell$ $tables$ contain cells with the same value merged into one cell.
\end{itemize}
\item $Nested$ $tables$ $contain$ $tables$ nested into cells.
\item $Multivalued$ $tables$ contain cells where there are multiple values.
\begin{itemize}
\item $Simple$ $multivalued$ $tables$ contain multivalued cells whose values belong to the same domain.
\item $Composed$ $multivalued$ $tables$ contain multivalued cells whose values belong to the different domains.
\end{itemize}
\end{enumerate}

\subsection{Header}
We classify table headers with respect to the following characteristics.
\begin{enumerate}
\item $No-header$ $tables$ refer to tables without any header.
\item $Simple$ $header$ $tables$ contain only one-level “usual” headers.
\item $Hierarchical$ $header$ $tables$ contain headers with multiple levels (Table 1). 
$Simple$ $header$ $tables$ and $Hierarchical$ $header$ $tables$ are subdivided with respect to header arrangement.

\begin{itemize}
\item Each header in $Single$ $header$ $tables$ is arranged in only one single row/column, without repetition.
\begin{figure}[h]
\begin{center}
 \includegraphics[width=12cm]{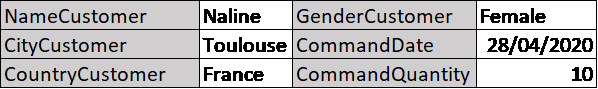}
 \caption{Distributed header table} \label{DistriHeader}
\end{center}
\end{figure}
\item The header of $Distributed$ $header$ $tables$ is distributed in different rows/columns (Table in FIG. 3).
\begin{figure}[h]
\begin{center}
 \includegraphics[width=12cm]{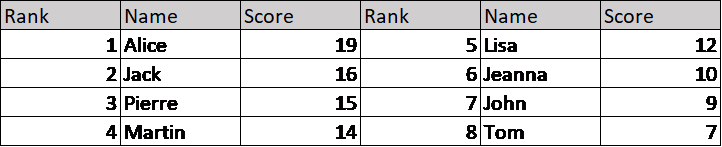}
 \caption{Duplicated header table} \label{DupliHeader}
\end{center}
\end{figure}
\item $Duplicated$ $header$ $tables$ have headers repeated many times (Table in FIG. 4).
\end{itemize}
\end{enumerate}

\citep{Zhang20} define elements of a table, including table page title, table caption, table headings, table cell, table row, table column and table entities. Table page title and table caption are explanatory elements of a table. Table entities concern the content and topic of a table. The other elements are all considered in our proposed classification. Moreover, our classification is based on the state of the art, all the possible cases mentioned in  existing tabular data typologies are taken into account. Thus, we believe that this typology is complete.

\section{Tabular data integration approach}

FIG. 5 illustrates our approach which aims to integrate automatically tabular data for the OLAP analysis. We first extract and identify tables (Section 4.1). The data source may contain one or several tables. If there is only one table, we transform the table type (Section 4.2), select the measures (Section 4.3), detect functional dependencies (Section 4.4) and finally generate the multidimensional schema (Section 4.5). If there are several tables, we either match the tables into one table by using by similarity measures and proceed as above, or generate a multidimensional schema for each table as above and match the schemas.

\begin{figure}[h]
\begin{center}
 \includegraphics[width=12cm]{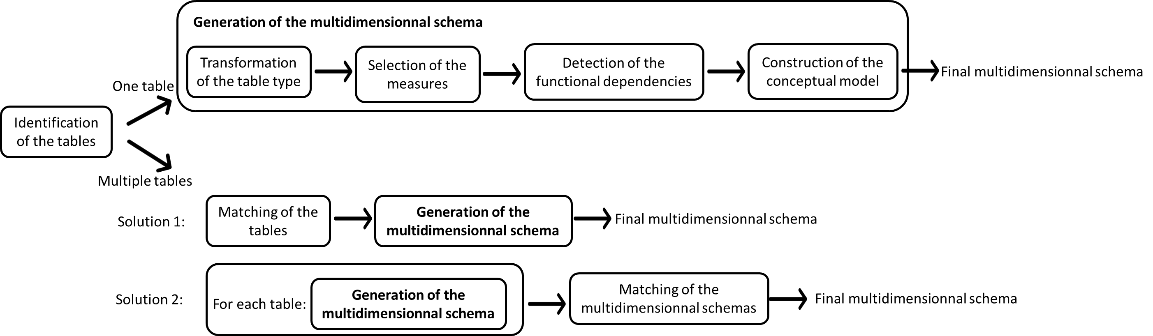}
 \caption{Overview of our tabular data integration approach} \label{OverviewProc}
\end{center}
\end{figure}

\subsection{Table identification}
Since the user we target are individuals, small enterprises or organizations without knowledge of BI, data sources may be of various types, e.g. CSV files, spreadsheet files, HTML tables or even images. There may also be several tables in some sources. Therefore, the first thing to do is to identify the sources and extract the tables within. Many table extraction methods are surveyed in \citep{Zhang20}, especially for web tables. There also exist deep learning methods to detect tables from images \citep{Schreiber17}; \citep{Paliwal20}. Let us first discuss our integration solution for the one-table case. We generalize it in Section 4.6.

\subsection{Table type transformation}
First, we should transform multidimensional source tables into a one-dimensional table. \citep{Milosevic16} propose an helpful method allowing the identification of multidimensional table structures. However, multidimensional tables may already contain obvious multidimensional structures that we can employ directly for the construction of a multidimensional schema. If this was the case, transforming the multidimensional table would be useless and time-wasting. In this article, we propose to perform the transformation to get a uniform representation of tables so that we can easily get the attributes and values for the following steps of our approach. We consider multidimensional tables  as a special case that we will process specifically.

For tables containing non-simple cells, it is necessary to convert these cells into simple cells by decomposing merged cells and multivalued cells by adding new cells or columns. Distributed and duplicated headers need to be detected by the algorithm of \citep{Yoshida01} in order to reform the table as a single-header table. In hierarchical header tables, hierarchical relations should be identified as candidates to hierarchies in the multidimensional model. Hierarchies can be extracted by the method proposed by \citep{Chen13}. Eventually, there may sometimes be even no attribute header in a table. In this case, we can use column identification methods \citep{Zhang20} to retrieve the semantic label of columns by matching them with knowledge bases.

\subsection{Measure selection}
To integrate a one-dimensional table for OLAP, we must firstly define measures. Measures are usually obtained by aggregation and calculation (max, min, count…), so most of them are numeric data. Normally, during the data warehouse design process, facts and measures are chosen by the user. As we aim to automate this process, we choose all numeric attributes in tables as candidate measures. However, not all numeric data can be considered as measures. For example, some numeric data may be IDs or Booleans. Yet, we can use the semantic labeling of numeric data to determine whether an attribute can be treated as a measure. \citep{Alobaid19} propose algorithms to identify the type of numeric data, including nominal data, ordinal data, intervals and ratios. Intervals and ratios are more likely to be measures. 

This automatic process can provide candidate measures to the users, but we may wrongly choose or omit some measures. For instance, a measure calculated by the count of an attribute does not need to be numeric. Calculating measures is also a big problem, because without knowing the exact user requirements, we cannot know the exact aggregation function. We can only propose basic functions such as count, sum, average, minimum and maximum, while measures may also be calculated by complex formulas on one or several attributes. Thence, if the user has specific requirements, we should also let her/him specify measures and correct the candidate measures we propose automatically. 

\subsection{Functional dependency detection}
The objective of this step is to determine the multidimensional model’s dimensions and associated hierarchies. To meet these needs, we apply the principles of functional dependencies between attributes. There exists a lot of algorithms to discover functional dependencies \citep{Liu12}; \citep{Papenbrock15}. For a relation schema $R$, $X$ and $Y$ represent a subset of the attributes of $R$. In a statement of functional dependency $X$ $\rightarrow$ $Y$, $X$ is usually called the left-hand side and $Y$ is called the right-hand side. We are only interested in the functional dependencies that have one attribute only in the right-hand side and get the minimal cover which is the minimal set of functional dependencies being able to infer all the functional dependencies. Hence, we obtain the elementary functional dependency such that no strict subset of the left-hand side determines the right-hand side. Moreover, we delete all the transitivity and pseudo-transitivity dependencies. We are also only interested in the functional dependencies whose left-hand side has one attribute only, since there is one attribute in each level of a hierarchy. 

To make sure that the functional dependencies that we discover conform to the dependency relationship of attributes in the real world, we hypothesize that there is enough data in terms of quantity and variety so as to represent real dependency relationships. Moreover, there should be no error in data nor empty data, but if this was the case, we could detect approximate functional dependencies. There are many methods for approximate functional dependency \citep{Liu12}, which propose different error measures. We can set a satisfaction threshold to decide the approximate functional dependencies. We can also search and complete empty data by querying open data \citep{Eberius12}. 

\subsection{Multidimensional schema detection}
\begin{sloppypar}
Given all the functional dependencies that satisfy our requirements, we can connect them to infer all hierarchies and even draw a functional dependency diagram to visualize them. For example, let us consider a product order table with attributes $idCustomer$, $nameCustomer$, $cityCustomer$, $countryCustomer$, $classCustomer$, $idProduct$, $nameProduct$, $categoryProduct$ and $quantity$. With quantity chosen as the measure, we get a fact $(F_1, {quantity})$. We may get the functional dependencies by applying the algorithm of functional dependency detection if the data of the table satisfy the hypothesis mentioned in the section 4.4: $idCustomer$ $\rightarrow$ $nameCustomer$, $nameCustomer$ $\rightarrow$ $cityCustomer$, $cityCustomer$ $\rightarrow$ $countryCustomer$, $nameCustomer$ $\rightarrow$  $classCustomer$, $idProduct$ $\rightarrow$ $nameProduct$, $nameProduct$ 	$\rightarrow$ $categoryProduct$. We can then obtain three hierarchies, but we cannot know whether an attribute is a parameter or a weak attribute, so we consider all the attributes as parameters. Thus, the three hierarchies are: $(H_1, <idCustomer, nameCustomer, cityCustomer, countryCustomer>), (H_2, < idCustomer, nameCustomer, classCustomer>), (H_3, <idProduct,$ $nameProduct,$ $categoryProduct>)$. The dependency graph is provided in FIG. 6:
\end{sloppypar}
\begin{figure}[h]
\begin{center}
 \includegraphics[width=12cm]{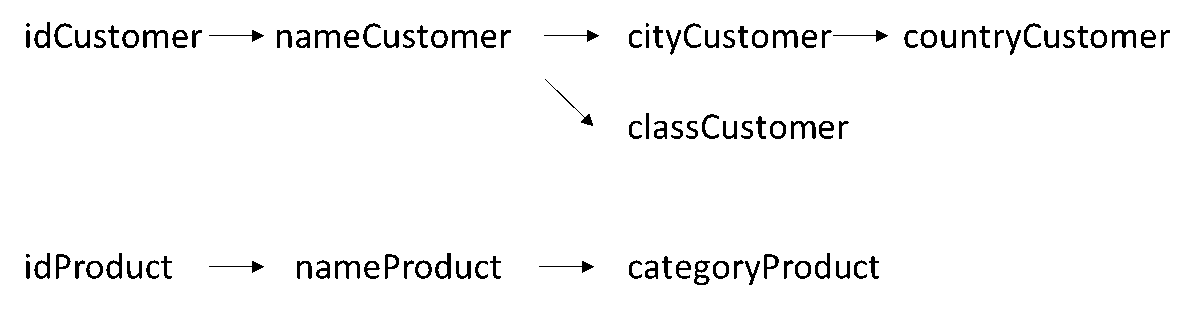}
 \caption{Dependency graph of the example} \label{DepenGraph}
\end{center}
\end{figure}
\begin{sloppypar}
All the hierarchies that share the same root have the same dimension, as we can see in FIG. 6, we can get 2 dimensions: $(D_1,$ $\{idCustomer,$ $nameCustomer,$ $cityCustomer,$ $countryCustomer,$ $classCustomer\},$ ${H_1, H_2})$ ,$ (D_2,$ $\{idProduct,$ $nameProduct,$ $categoryProduct\},$ ${H_3})$
\end{sloppypar}
The final step of our approach is to simply link measures to dimensions. We do not need to verify whether measures are dependent on the totality of the root parameters, because measures are obtained by the aggregation of attributes. If we aggregate all root parameters, measures are certainly dependent on the totality of the root parameters, since there is no redundancy after aggregation. Therefore, we obtain a conceptual multidimensional model $(nameSchema, F_1, \{D_1, D_2\})$.

If there are equivalent attributes, $i.e. X, Y$ are subsets of a relation schema $R$, $X$ and $Y$ are equivalent attributes if $X$$\rightarrow$$Y$ and $Y$$\rightarrow$$X$, there is probably a parameter and its weak attributes inside an equivalent attribute set, so we should know what attribute may be an ID so that it can be considered as a parameter. We could also introduce semantic solutions for solving these problems. 

\subsection{Integration of multiple source tables}
Our approach currently applies to the one-table case. In case of multiple tables, we propose two potential matching solutions, by using different schema matching and data matching technologies with the help of the similarity measures. We can either match all tables into one single table and generate automatically the multidimensional model of the table, or we can generate automatically the multidimensional model of each table and match these schemas to one single schema. The fusion can also be described by a manipulation algebra \citep{Ravat05}. 

\section{Implementation}

Since we are still studying the integration of multisource tables, we have only implemented the automatic multidimensional schema generation algorithm for one table source in Python. To illustrate our approach, we test with a csv file from the open data site of the French government\footnote{https://www.data.gouv.fr/fr/datasets/temps-de-parole-des-hommes-et-des-femmes-a-la-television-et-a-la-radio/} . This is a file recording speaking time of men and women corresponding to more than a million hours of programs broadcast from 1995 to February 28, 2019.

\begin{figure}[h]
\begin{center}
 \includegraphics[width=12cm]{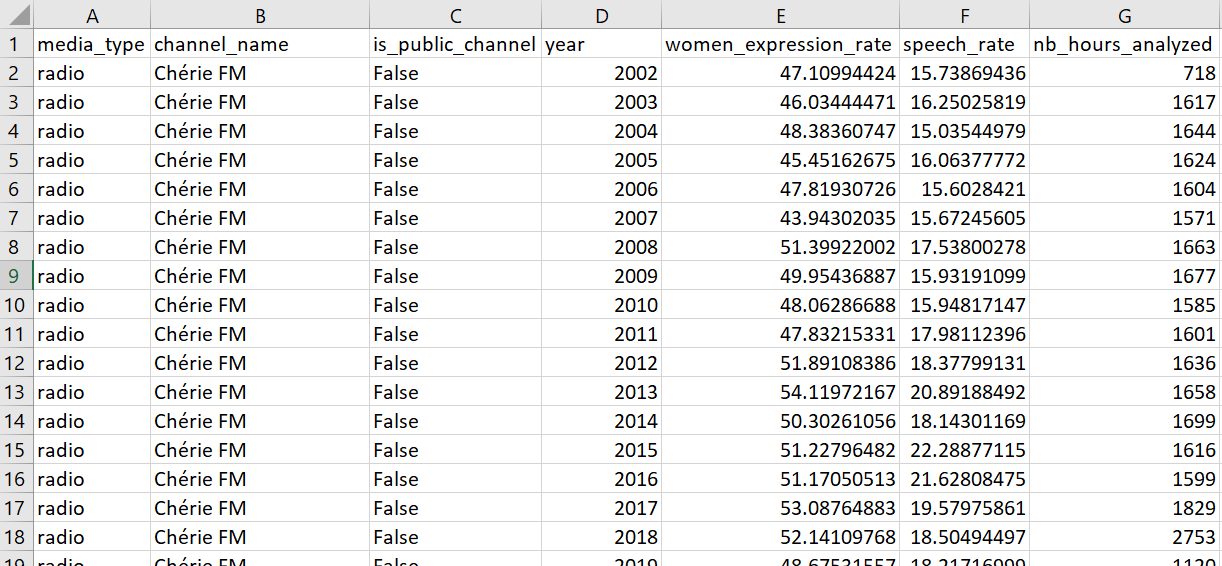}
 \caption{Test file excerpt} \label{TestFile}
\end{center}
\end{figure}

As illustrated in FIG. 7, the file contains 7 columns: media type, channel name, is public channel, year, women expression rate, speech rate and number of the analyzed hours. There are 21 radio stations and 34 TV channels recorded in the file. The result of our algorithm is shown in FIG. 8.

\begin{figure}[h]
\begin{center}
 \includegraphics[width=12cm]{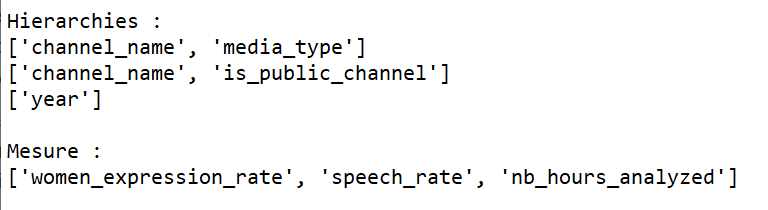}
 \caption{Test file excerpt} \label{Result}
\end{center}
\end{figure}

We detect 3 measures: women expression rate, speech rate and number of the analyzed hours. We get 3 hierarchies, since there are 2 hierarchies sharing the same root, they are in the same dimension. So, we get 2 dimensions. 

The user may also need the men expression rate, she/he can just add this measure by a simple calculation (1 – women expression rate). The user can also give a name to the dimensions, then we can get the multidimensional schema (FIG. 9).

\begin{figure}[h]
\begin{center}
 \includegraphics[width=12cm]{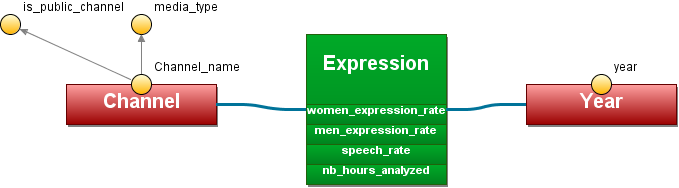}
 \caption{Final multidimensional schema} \label{MultiSchema}
\end{center}
\end{figure}

\section{Conclusion and future research}
In this paper, we study how to automate the integration of tabular data to generate multidimensional data warehouses. First, we propose a taxonomy of tabular data to help in the table identification step. Then, our table identification solution consists in identifying the table, transforming it into a one-dimensional table and generating the multidimensional schema by exploiting functional dependencies. Matching the tables then generating a multidimensional schema or generating a multidimensional schema for each table and then matching the schemas are two possible solutions for the integration of multiple tables. 

In ongoing work, we will implement and refine the idea that we propose. We will test the two possible methods for the generalization of multiple tables and evaluate their feasibility and performance to decide the one to use. Once we succeed in generating automatically the conceptual multidimensional schema, we will define an automatic process to populate the schema from the data and the content of the source table.

\section*{Acknowledgements}

The research depicted in this paper is funded by the French National Research Agency (ANR), project ANR-19-CE23-0005 BI4people (Business Intelligence for the people).

\bibliographystyle{rnti}
\bibliography{biblio}

\Eng

\end{document}